\documentclass[twocolumn,showpacs,preprintnumbers,amsmath,amssymb]{revtex4-1}

\usepackage{epsfig}
\usepackage{graphicx}
\usepackage{dcolumn}
\usepackage{bm} 

\newcommand{\beq}{\begin{equation}}
\newcommand{\eeq}{\end{equation}}
\newcommand{\beqa}{\begin{eqnarray}}
\newcommand{\eeqa}{\end{eqnarray}}

\newcommand{\ket}[1]{|#1\rangle}                
\newcommand{\bra}[1]{\langle#1|}                
\newcommand{\matel}[3]{\langle#1|#2|#3\rangle}  

\def\pra#1{{ Phys.\ Rev. A\/} {\bf#1}}
\def\prb#1{{ Phys.\ Rev. B\/} {\bf#1}}

\def\prl#1{{ Phys.\ Rev.\ Lett.} {\bf#1}}

\begin{document}

\title{Witnessing non-classicality of a quantum oscillator state by coupling it to a qubit}

\author{S. Agarwal and J.H. Eberly}

\affiliation{ Rochester Theory Center and the Department of Physics
\& Astronomy\\
University of Rochester, Rochester, New York 14627}

\email{shantanu@pas.rochester.edu}


\date{\today}

\begin{abstract}
We propose a new witness operation for the non-classical character of a harmonic oscillator state. The method does not require state reconstruction. For all harmonic oscillator states that are classical, a bound is established for the evolution of a qubit which is coupled to the oscillator. Any violation of the bound can be rigorously attributed to the non-classical character of the initial oscillator state. 
\end{abstract}

\pacs{03.65.Wj, 42.50.Dv, 03.67.-a}


\maketitle

\section{Introduction}
Quantum mechanics is fundamentally different from classical mechanics. Interference of probability amplitudes, superposition of states, uncertainty relations between canonically conjugate variables, etc., are essential quantum phenomena that are not present within the classical theory. If a state of a system exhibits any such intrinsically quantum feature, the state is called non-classical \cite{Mandel-86}. 

For a given physical system, quantitatively categorizing the states into classical and non-classical is usually challenging. In this report, we are interested in categorizing the states of a harmonic oscillator. Any harmonic oscillator state can be written in the coherent state diagonal representation \cite{Sudarshan-63,Glauber-63}:
\begin{align}
\rho=\int\mathrm{d^2}\alpha\, P(\alpha)\ket{\alpha}\bra{\alpha},
\end{align}
where $\ket{\alpha}$ is a coherent state. Within the fundamental limits imposed by the uncertainty relation between the position and the momentum, a coherent state corresponds as closely as possible to a classical harmonic oscillator of a definite complex amplitude. For this reason, a coherent state can be considered to be classical. A natural definition of classicality based on this observation was introduced by Glauber \cite{Glauber-63}. He proposed that if $P(\alpha)$ is a valid probability measure, the state $\rho$ can be thought of as a statistical mixture of various classical states and thus is classical itself. On the other hand, if $P(\alpha)$ is not a valid probability measure, the state $\rho$ is non-classical. We will adopt this definition of non-classicality of a harmonic oscillator state.

There are many ways of checking whether the $P$ distribution corresponding to a given oscillator state fails to be a valid probability measure. For example, one can perform a complete state tomography to find out what $\rho$ is \cite{Leonhardt-95, Lvovsky-09}. Knowing the state, one can then derive the $P$ distribution. Such an approach, although possible in principle, is difficult to carry out in practice because the inverse relation connecting the distribution to the state involves an integral with an exponentially diverging factor \cite{Mehta-67}:
\begin{align}
P(\alpha)=\frac{e^{|\alpha|^2}}{\pi^2}\int\mathrm{d^2}\gamma\,e^{|\gamma|^2}\matel{-\gamma}{\rho}{\gamma}e^{-\gamma\alpha^*+\gamma^{*}\alpha}.
\end{align} 
Because of the diverging term, any experimental error gets exponentially enhanced \cite{Meas-Uncer}. For this reason, instead of performing a full state tomography, one generally looks for non-classicality witnesses that signify $P$ as an invalid probability measure. Some of these witnesses are anti-bunching \cite{Kimble-76}, sub-Poissonian statistics \cite{Mandel-79, Short-84}, quadrature squeezing \cite{Slusher-85}, slower decay of the characteristic function of the rotated quadrature distribution than the characteristic function of the ground state \cite{Vogel-00}, negative Wigner distribution of the oscillator state \cite{Lvovsky-09}, violation of Bochner's criterion for the existence of a valid positive semidefinite characteristic function of the $P$ distribution \cite{Vogel-00, Mari-11}, etc.

\section{Witnessing non-classicality through a qubit coupled to the oscillator}

In this report, we propose another observable signature of non-classicality of an oscillator state. We longitudinally couple the oscillator to a qubit with the joint Hamiltonian given by:
\begin{align}\label{e.Ham}
H=\hbar\frac{\omega_0}{2}\sigma_{z}+\hbar\omega a^{\dagger}a+\hbar\omega\beta(a+a^{\dagger})\sigma_{z},
\end{align}
where $\omega_0$ and $\omega$ are the qubit and oscillator frequencies and $\beta$ is a dimensionless, constant coupling parameter. The oscillator operators, $a$ and $a^{\dagger}$, are the annihilation and creation operators respectively and the qubit operator, $\sigma_z$, is a Pauli matrix. This Hamiltonian is different from the Rabi Hamiltonian where the qubit-oscillator coupling is through $\sigma_x$.  

Because of the qubit-oscillator interaction, it is possible to learn about the initial state of the oscillator by following the evolution of the qubit. The method of reconstructing an arbitrary oscillator state by looking into the dynamics of an interacting few level system was proposed in \cite{Sascha-95}. Later, using an interacting qubit, the method of state reconstruction was developed in areas of cavity QED \cite{Lutterbach-96, Bodendorf-98, Kim-98, Nogues-00, Bertet-02, Amaro-03, Zou-04, Deleglise-08}, circuit QED \cite{Melo-06, Hofheinz-09}, trapped ions \cite{Zheng-05, Zhen-06} and nano-mechanical resonators \cite{Singh-10, Tufarelli-11}. 

It was shown by Tufarelli, et al., that by using a Hamiltonian that is similar to Eq. (\ref{e.Ham}), one can reconstruct the entire state of the oscillator \cite{Tufarelli-11} . This state reconstruction method is experimentally extremely challenging as it requires a predefined, deterministic modulation of the coupling strength in time. Also in the reconstruction procedure, it is necessary to be able to change the amplitude of modulation in different experimental runs. In the present report, we propose a different method of witnessing the oscillator state non-classicality through the qubit dynamics that can be carried out using a constant coupling parameter, $\beta$. 

We assume that the qubit and the oscillator states are initially separable and their joint state is given by:
\begin{align}
\rho_{qo}&=\rho_{q}\otimes\rho_{o},\nonumber\\
&=\begin{pmatrix}
z(0) & w(0)\\
w^*(0) & 1-z(0)\\
\end{pmatrix}
\otimes \int\mathrm{d^2}\alpha\, P(\alpha)\ket{\alpha}\bra{\alpha}.
\end{align} 
The oscillator state, $\rho_{o}$, is the state whose non-classicality we want to investigate. The rows and columns of the qubit state are defined in the basis: $\ket{+}$ and $\ket{-}$, where $\sigma_z\ket{\pm}=\pm\ket{\pm}$. The parameters, $z$ and $w$, defining the matrix elements of the qubit are related to the expectation values of the Pauli matrices: $z=(\langle\sigma_z\rangle+1)/2$, $w=(\langle{\sigma_x}\rangle-i\langle{\sigma_y}\rangle)/2$.

Using the Hamiltonian, Eq. (\ref{e.Ham}), the time evolution of the joint state, $\rho_{qo}$, can be found. The qubit state can be evaluated by tracing over the oscillator degrees of freedom from the time evolved joint qubit-oscillator state:
\begin{align}
\rho_{q}(t)=\mathrm{Tr_{o}}\{e^{-iHt/\hbar}\rho_{qo}e^{iHt/\hbar}\}.
\end{align}
The diagonal terms of the qubit density matrix, written in the $\sigma_z$ eigenbasis, do not change in time, i.e. $z(t)=z(0)$. This is because $\sigma_{z}$ is a constant of motion: $\left[\sigma_z,H\right]=0$. For the off-diagonal terms, we find:
\begin{align}\label{e.def_gt}
w(t)&=e^{-i\omega_0 t}g(t)w(0),
\end{align}
where $g(t)$ depends only on the oscillator degrees of freedom and is defined as 
\begin{align}
g(t)=&\mathrm{Tr_{o}}\{e^{-iH_{+}t/\hbar}\rho_{o}e^{iH_{-}t/\hbar}\},\nonumber\\
=&\int\mathrm{d^2}\alpha\,P(\alpha)\nonumber\\
&\times\mathrm{Tr_{o}}\{e^{-iH_{+}t/\hbar}\ket{\alpha}\bra{\alpha}e^{iH_{-}t/\hbar}\},
\end{align}
where 
\begin{align}
H_{\pm}=\hbar\omega\left(a^{\dagger}a\pm\beta(a+a^{\dagger})\right).
\end{align}
The $H_{\pm}$ operators correspond to the Hamiltonians of displaced harmonic oscillators. So, one can evaluate $g(t)$ analytically to get:
\begin{align}
g(t)&= e^{-8\beta^2\sin^2\left(\frac{\omega t}{2}\right)}\int\mathrm{d^2}\alpha\,P(\alpha)\nonumber\\ 
&\quad\times e^{-4i\beta(\alpha e^{-i\omega t/2}+\alpha^* e^{i\omega t/2})\sin\frac{\omega t}{2}},\nonumber\\
&= e^{-8\beta^2\sin^2\left(\frac{\omega t}{2}\right)}\int\mathrm{d^2}\alpha\,P(\alpha)\, f(\alpha,t),\label{e.gt}\\
&= e^{-8\beta^2\sin^2\left(\frac{\omega t}{2}\right)}W(t),
\end{align}
where we have defined
\begin{align}
f(\alpha,t)&=e^{-4i\beta(\alpha e^{-i\omega t/2}+\alpha^* e^{i\omega t/2})\sin\frac{\omega t}{2}},
\end{align}
and
\begin{align}
W(t)&=\int\mathrm{d^2}\alpha\,P(\alpha)\, f(\alpha,t).
\end{align}

If we look at the absolute value of the function $g(t)$, we get from Eq. (\ref{e.gt}):
\begin{align}\label{e.g_ineq}
|g(t)|&=e^{-8\beta^2\sin^2\left(\frac{\omega t}{2}\right)}\Big|\int\mathrm{d^2}\alpha\,P(\alpha)\, f(\alpha,t)\Big|,\nonumber\\
&\leq e^{-8\beta^2\sin^2\left(\frac{\omega t}{2}\right)}\int\mathrm{d^2}\alpha\,\Big|P(\alpha)\, f(\alpha,t)\Big|.
\end{align}
If $P(\alpha)$ is a valid probability measure, we can write
\begin{align}\label{e.P_ProbMes}
\Big|P(\alpha)\, f(\alpha,t)\Big|&=P(\alpha) \Big|f(\alpha,t)\Big|.
\end{align}
Putting Eq. (\ref{e.P_ProbMes}) in Eq. (\ref{e.g_ineq}) and using the fact that $|f(\alpha,t)|=1$ and $\int\mathrm{d^2}\alpha\,P(\alpha)=1$, we get:
\begin{align}
|g(t)|&\leq e^{-8\beta^2\sin^2\left(\frac{\omega t}{2}\right)}\int\mathrm{d^2}\alpha\,P(\alpha), \nonumber\\
&=e^{-8\beta^2\sin^2\left(\frac{\omega t}{2}\right)}.
\end{align}
Using this upper bound for $g(t)$ in the expression for the off diagonal element of the qubit density matrix, Eq. (\ref{e.def_gt}), we get:
\begin{align}\label{e.uv_inq}
|W(t)| \equiv e^{8\beta^2\sin^2\left(\frac{\omega t}{2}\right)}\Big|\frac{w(t)}{w(0)}\Big| \leq 1.
\end{align}
Note that $w(t)$ and thus $|W(t)|$ can be experimentally measured by measuring the expectation value of the qubit observables, $\sigma_x$ and $\sigma_y$, as a function of time. 

Inequality (\ref{e.uv_inq}) is the main result of the paper and we call it the non-classicality witness inequality (NCWI). The inequality states that if the oscillator state is such that the associated $P$ distribution is a proper probability measure implying Eq. (\ref{e.P_ProbMes}) to be correct, $|W(t)|$ is always bounded from above by unity. Thus if $|W(t)|>1$, we know for sure that the initial oscillator state, $\rho_0$, is non-classical.  


Two remarks are in order at this point. First, it can be shown that $W(t)$ is related to the characteristic function of the $P$ distribution \cite{Agarwal-11}. This suggests that Eq. (\ref{e.uv_inq}) is related to the criterion for observing non-classicality through measuring the probability distribution of rotated quadratures \cite{Vogel-00}. The second remark is just that a violation of the NCWI is only a sufficient but not a necessary condition for determining whether a state is non-classical, i.e., a state can be non-classical and still not violate the witness inequality (see the examples of Fock states, vacuum subtracted thermal states and Schr\"{o}dinger Cat states in the next section).

\section{Examples}
To illustrate the above statements, we calculate $|W(t)|$ for various initial oscillator states. 

\vspace{0.25cm}
\textit{Coherent state:} The $P$ distribution for a coherent state, say $\ket{\alpha_0}$, is $P_{coh}(\alpha)=\delta^{2}(\alpha-\alpha_0)$. Corresponding to $P_{coh}(\alpha)$, we have $|W_{coh}(t)|=1$. Thus, no coherent state ever violates the NCWI. This is required for consistency because by our definition a coherent state is considered to be a classical state. 
\vspace{0.25cm}

\textit{Thermal state:} For a thermal state with mean excitation number $\bar{n}$, the $P$ distribution is given by:
\begin{align}
P_{th}(\alpha)=\frac{1}{\pi\bar{n}}\exp{(-|\alpha|^{2}/\bar{n})}.
\end{align}
We see that $P_{th}(\alpha)$ is a Gaussian probability distribution and thus a thermal state with arbitrary $\bar{n}$ can be considered classical. Using $P_{th}(\alpha)$, we get
\begin{align}
|W_{th}(t)|=e^{-16\bar{n}\beta^2\sin^2\left(\frac{\omega t}{2}\right)}.
\end{align}
In agreement with our intuition that a thermal state is a classical state, we see that $|W_{th}(t)|$ is always less than one and thus never violates the NCWI.
\vspace{0.25cm}

\textit{Fock state:} The $P$ distribution for a Fock state, $\ket{N}$, is
\begin{align}
P_{N}(\alpha)=\frac{\exp{(\alpha\alpha^{*})}}{N!}\left(\frac{\partial^{2N}}{\partial\alpha^{N}\partial\alpha^{*N}}\delta^{2}(\alpha)\right).
\end{align} 
Because $P_{N}(\alpha)$ for $N>0$ involves derivatives of the delta function and is more singular than the delta function itself, each Fock state, except the ground state, is non-classical. Calculating $W_{N}(t)$ using $P_{N}(\alpha)$, we get:
\begin{align}
|W_{N}(t)|=\Big|L_{N}\left(16\beta^2\sin^2{\frac{\omega t}{2}}\right)\Big|,
\end{align}
where $L_{N}(x)$ is a Laguerre polynomial. 

\begin{figure}[t]

\includegraphics[width=8.5cm, height=5cm]{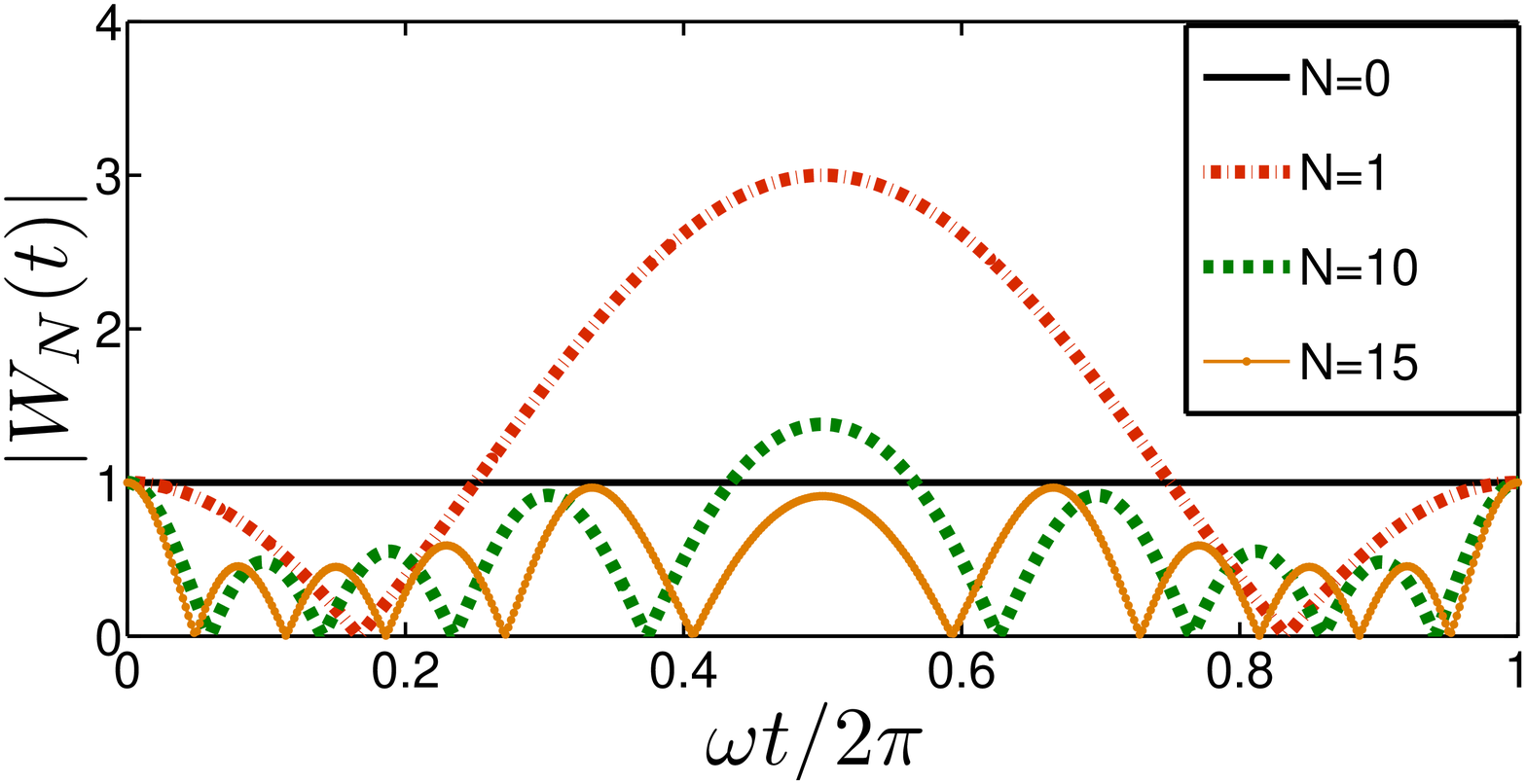}
\caption{(Color online) The function $|W_N(t)|$ for $\beta=0.5$ and various values of $N$. We clearly see that $|W_N(t)|$ violates the NCWI indicating that Fock states are non-classical states.}\label{fig:W_N}
\end{figure}

In Fig. \ref{fig:W_N}, we plot $|W_{N}(t)|$ for various values of $N$ and for a given constant coupling strength, $\beta=0.5$. Noting that $|W_{N}(t)|$ is a periodic function, we plot $|W_{N}(t)|$ only for a single period. 
We see from the figure that for some Fock states, e.g. for $N=1$ and $10$, the NCWI is violated. This agrees with the non-classical nature of these Fock states. On the other hand, it is important to note that for any given coupling strength, not all Fock states will violate the NCWI. For example, in Fig.  \ref{fig:W_N}, no violation is shown by $|W_{0}(t)|$ and $|W_{15}(t)|$. However, this result does not suggest that some Fock states are classical. This point is further examined in the following examples of vacuum subtracted thermal state and superposition of classical states. By looking at the asymptotic expression of a Laguerre polynomial for large $N$ \cite{Abramowitz}, it can be shown that for a given coupling strength, $\beta$, the NCWI will not be violated for highly excited Fock states. 


\vspace{0.25cm}

\textit{Vacuum subtracted thermal state:} We now consider the state
\begin{align}
\rho'_{th}=\sum_{N=1}^{\infty}2^{-N}\ket{N}\bra{N}.
\end{align}
The above state corresponds to a thermal state with unit mean excitation number and from which the vacuum state has been discarded. The $P$ distribution for this state is
\begin{align}
P'_{th}(\alpha)=\frac{2}{\pi}\exp{(-|\alpha|^{2})}-\delta^2(\alpha).
\end{align}
We see that $P^{'}_{th}(\alpha)$ has a negative measure at $\alpha=0$ \cite{Glauber-63}. This implies that $\rho'_{th}$ is a non-classical state. Corresponding to $P^{'}_{th}(\alpha)$, we have
\begin{align}
|W'_{th}(t)|=|2e^{-16\beta^2\sin^2\left(\frac{\omega t}{2}\right)}-1|.
\end{align}
Although $P^{'}_{th}(\alpha)$ corresponds to a non-classical state, we see that $|W'_{th}(t)|$ is always less than one and never violates the witness inequality. This illustrates the fact that violation of the NCWI is not a necessary but only a sufficient condition for determining non-classicality \cite{Diosi-00}.    
\vspace{0.25cm}

\textit{Superposition of classical states:} Even though a coherent state is considered to be classical, a state consisting of a superposition of coherent states might exhibit non-classicality. To understand this, let us examine the Schr\"{o}dinger Cat state which consists of an equal superposition of two coherent states, one being the negative of the other:
\begin{align}
\ket{\Psi_{sc}(\alpha_0)}=\frac{\ket{\alpha_0}+\ket{-\alpha_0}}{\sqrt{2(1+e^{-2\alpha_0^2})}}.
\end{align} 
For simplicity, we have taken $\alpha_0$ to be real. The $P$ distribution corresponding to $\ket{\Psi_{sc}(\alpha_0)}$ is \cite{Mandel-86}:
\begin{align}
P_{sc}(\alpha)&=\mathcal{N}^2\Big[\delta^{2}(\alpha-\alpha_0)+\delta^{2}(\alpha+\alpha_0)+e^{\left(|\alpha|^2-|\alpha_0|^2\right)}\nonumber\\
& \times\Big(e^{\alpha_0\partial_{\alpha^*}}e^{-\alpha_0\partial_{\alpha}}+e^{-\alpha_0\partial_{\alpha^*}}e^{\alpha_0\partial_{\alpha}}\Big)\delta^{2}(\alpha)\Big],
\end{align}
where $1/\mathcal{N}^2=2(1+e^{-2\alpha_0^2})$. We see that for $\alpha_0\neq0$, $P_{sc}(\alpha)$ contains infinitely high order derivatives of the delta function. This clearly suggests that a Schr\"{o}dinger Cat state is a non-classical state if $\alpha_0\neq0$. 

\begin{figure}[!t]
\includegraphics[width=8.5cm, height=5cm]{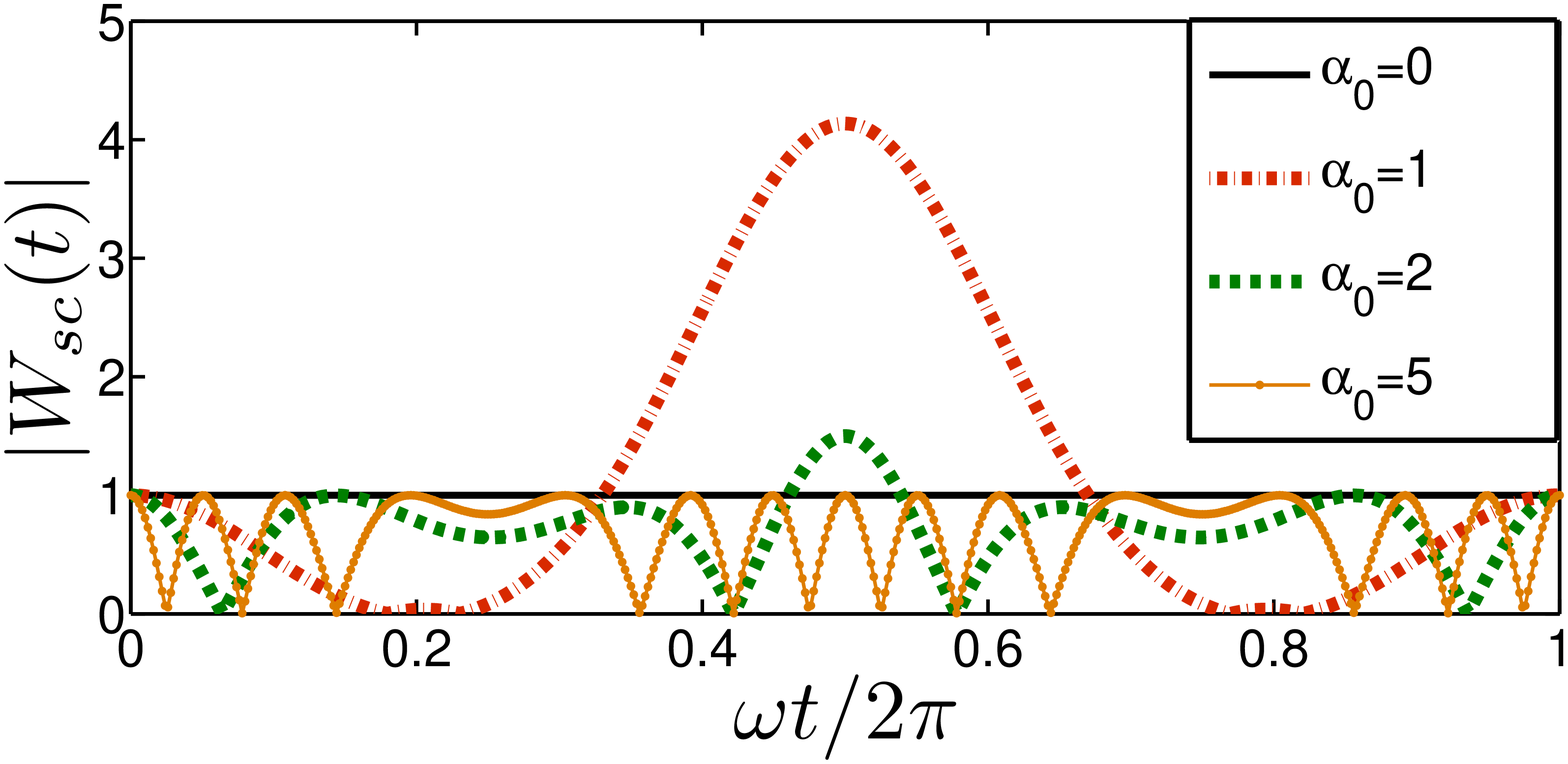}
\caption{(Color online) The function $|W_{sc}(t)|$ for $\beta=0.5$ and various values of $\alpha_0$. For some values of $\alpha_0$, $|W_{sc}(t)|$ violates the NCWI. This indicates the non-classical nature of these Schr\"{o}dinger cat states.}\label{fig:W_sc}
\end{figure}

For the Schr\"{o}dinger Cat state, $W_{sc}(t)$ can be calculated analytically to get:
\begin{align}
|W_{sc}(t)|&=\mathcal{N}^2\Big|\cos{(4\beta\alpha_0\sin\omega t)}\nonumber\\
&\quad+e^{-2\alpha^2}\cosh{\left(8\beta\alpha_0\sin^2\frac{\omega t}{2}\right)}\Big|.
\end{align}
In Fig. \ref{fig:W_sc}, we plot $W_{sc}(t)$ for $\beta=0.5$ and for various values of $\alpha_0$. It is clear that the NCWI is violated for $\alpha_0=1$ and $2$. This confirms the non-classical nature of these states. On the other hand, for any given value of $\beta$, we notice that not all Schr\"{o}dinger Cat states violate the NCWI. This is evident from the plots corresponding to $\alpha_0=0$ and $5$. 

\section{Concluding remarks}

In this report, we use Glauber's criterion for non-classicality, which is the failure of an oscillator's coherent state $P$ distribution to behave as a valid classical probability distribution. We demonstrate a witness operation for non-classicality of this type. We show that one can imprint a signature of non-classicality onto the dynamics of an interacting qubit. The result is that all classical initial states of the oscillator lead to a fixed qubit bound, given in equation (\ref{e.uv_inq}). In other words, the qubit operator's time evolution, expressed via $|W(t)|$, must remain bounded from above if the oscillator's initial state was classical. A violation of the bound is a direct indication of the non-classical nature of the initial oscillator state. The method does not require state reconstruction, and monitoring of the qubit can be confined to a single period of the oscillator. A number of examples are presented to illustrate the behavior of the bound, and confirm intuitive expectation in special cases. This strategy for witnessing non-classicality is well suited for physical systems where strong longitudinal coupling can be achieved between a qubit and a single-mode of an oscillator \cite{Xue-07, Wilson-10}. Because of the unavoidable interaction of the qubit-oscillator system with its environment in any experiment, it is important to take into account the effect of noises on the evolution of $|W(t)|$. Under the Markovian approximation of system-environment interaction, one expects $|W(t)|$ to decrease exponentially in time and this decay can be compensated for if it is experimentally possible to obtain the exponential decay constants \cite{Tufarelli-11}.

\begin{acknowledgments}
Financial support was received from NSF PHY-1203931 and ARO W911NF-09-1-0385.
\end{acknowledgments}


\begin{thebibliography}{99}
\bibitem{Meas-Uncer} This point is discussed in section 2.4 of Ref. \cite{Leonhardt-95}.
\end{thebibliography}
\end{document}